\documentclass[sigconf]{acmart}
\renewcommand\footnotetextcopyrightpermission[1]{}
\usepackage{todonotes}
\usepackage{soul}
\usepackage[utf8]{inputenc}
\usepackage{float}

\def\BibTeX{{\rm B\kern-.05em{\sc i\kern-.025em b}\kern-.08emT\kern-.1667em\lower.7ex\hbox{E}\kern-.125emX}}
    
\begin{document}

% \title{Towards More Robust Online Safety: Uncovering Evasion Vulnerabilities in AI Hate Speech Detectors}
\title{All You Need is "Leet": Evading Hate-speech Detection AI}
% All You Need is "Love": Evading Hate-speech Detection

\author{Sampanna Yashwant Kahu}
\authornote{Both authors contributed equally to this research.}
\orcid{0000-0002-8522-2926}
\email{sampanna@vt.edu}
\affiliation{
  \institution{Virginia Tech}
  \city{Blacksburg}
  \state{Virginia}
  \postcode{24060}
}

\author{Naman Ahuja}
\authornotemark[1]
\email{namanahuja@vt.edu}
\affiliation{
  \institution{Virginia Tech}
  \city{Blacksburg}
  \state{Virginia}
  \postcode{24060}
}

\renewcommand{\shortauthors}{Ahuja and Kahu.}

\begin{abstract}
Social media and online forums are increasingly becoming popular. Unfortunately, these platforms are being used for spreading hate speech. In this paper, we design black-box techniques to protect users from hate-speech on online platforms by generating perturbations that can fool state of the art deep learning based hate speech detection models thereby decreasing their efficiency. We also ensure a minimal change in the original meaning of hate-speech. Our best perturbation attack is successfully able to evade hate-speech detection for 86.8 \% of hateful text. \\

The source code and data used in this work is available at: https://github.com/SampannaKahu/all\_you\_need\_is\_leet.
\end{abstract}




\maketitle

\section{Introduction}
Hate speech has been rampant on the internet recently. Such harmful texts expose children and even adults to unwanted and unsafe content, and may also lead to polarization of opinions to cause conflicts. Considering the scale of the internet and social media platforms today, it is very difficult to enforce legislation in the virtual world. Thus, the need of the hour is to come up with ways to suppress this plague. With the advancements in computational power, many companies are actively working to create state of the art deep learning models to detect hate speech.  Microsoft offers Content Moderator\cite{content_moderator}, a machine-assisted content moderation API for images, text, and videos. Facebook\cite{fb_hate} in 2019 at its annual tech conference F8 claimed that they have made detection of hate content faster by using self-supervised learning. Also, it has recently banned various individuals cited for hate speech at its platform. The Perspective API\cite{perspective_api_website} from Jigsaw (a part of Google's parent company Alphabet) gives online comment moderators an evolving set of tools to combat abuse and harassment. Some of these models are provided as Machine Learning-as-a-Service (MLaaS). Generally, the model is deployed on the cloud servers, and users can only access the model via an API. Note that the free usage of the API might be limited among these platforms. Though deep neural network models have exhibited state-of-the-art performance in a lot of applications, recently they have been found to be vulnerable against adversarial examples which are carefully generated by adding small perturbations to the legitimate inputs to fool the targeted models\cite{Cheng2018Seq2SickET}\cite{Eykholt2017RobustPA}. The power of deep learning methods cannot be denied, but applications of such adversaries raise serious concerns. Earlier works\cite{article} have shown that even if the attacker has only a black box access to the model via an API, that is, the attacker is not aware of the model architecture, parameters or training data, and is only capable of querying the target model with output as the prediction or confidence scores, it is possible to affect the model outputs through adversarial inputs. The aim of this research project is to design black-box techniques to protect users from hate-speech on online platforms by generating perturbations that can fool state of the art deep learning based hate speech detection models, hence decreasing their efficiency. We also want to ensure minimum change in the original meaning of hate-speech. Thus, we measure the change this perturbation brings to the original text. After explaining and evaluating the performance of perturbation attacks, we propose some methods to defend against such attacks.

\subsection{Motivation}
The increasing popularity of social media platforms like Youtube, Facebook and Twitter have revolutionized communication, content sharing and advertisement. But, the anonymity offered by these platforms has led to an exponential increase in hate speech propagation on these platforms. American Bar Association defines hate speech as a speech that offends or insults groups based on race, colour, religion, national origin, sexual orientation, disability, or other traits. They are words that are hurtful, emotionally harmful, and psychologically stunning. Statistics show that in the US, hate speech and hate crime is on the rise especially since the Trump election\cite{HST}. As a matter of fact, the German government had threatened to fine social networks up to 50 million euros per year if they continue to fail to act on hateful postings\cite{gamback-sikdar-2017-using}. Recent surveys have shown that hate speech has become an almost unavoidable fact of life on the internet. More than half of Americans (53 percent) say they were subjected to hateful speech and harassment in 2018\cite{USAToday}. Threats online can spill over into real-world violence and turn deadly. Robert Bowers, who allegedly killed 11 people at a Pittsburgh synagogue in 2018, regularly posted anti-Semitic and neo-Nazi propaganda on Gab, a social network frequented by right-wing extremists. Cesar Sayoc, who's accused of mailing homemade explosive devices last year to critics of President Donald Trump, made repeated threats against public figures on Twitter\cite{USAToday}. The millions of hateful posts and videos polluting their platforms represent one of the most pressing challenges for Facebook, Twitter, YouTube and other technology companies. Measures such as hiring thousands of moderators and training artificial intelligence software to root out online hate and abuse have not yet solved the problem. All these instances tell us how important it is to eradicate the problem of hate on online platforms. The gravity of the matter can be judged by the plethora of international initiatives that have been launched towards the qualification of the problem and the development of counter-measures\cite{book}.
\subsection{Literature Survey}
Existing works on adversarial examples mainly focus on the image domain, generation of text-based adversarial samples being a relatively newer domain. Perturbation in the images can often be made virtually imperceptible to humans, causing both humans and state-of-the-art models to disagree. However, in the text domain, small perturbations might be clearly perceptible, with the replacement of a single word drastically altering the semantics of the sentence. Thus, in general, existing attack algorithms designed for images cannot be directly applied to text. Gröndahl et al. studied\cite{love} five model architectures presented in four papers to set up an experimental comparative analysis of state-of-the-art hate speech detection models and datasets (Wikipedia and Twitter). They also presented several attacks: word changes, word-boundary changes, and appending unrelated innocuous words which proved to be effective against all models. Hosseini et al.\cite{Hosseini2017DeceivingGP} demonstrated the vulnerability of Google's Perspective system against the adversarial examples. Through different experiments, they show that an adversary can deceive the system by misspelling the abusive words or by adding punctuation between the letters. They also proposed some
countermeasures to the proposed attack. But, when we checked the toxicity of their perturbed text via Perspective API, it now returns a high toxicity score, making their attacks futile. Li et al.\cite{TextBugger} proposed a framework that can effectively generate utility-preserving (i.e., keep its original meaning for human readers) adversarial texts against state-of-the-art text classification systems under both white-box and black-box settings. In the white-box scenario, they first find important words by computing the Jacobian matrix of the classifier and then choose an optimal perturbation from the generated five kinds of perturbations. In the black-box scenario, they first find the important sentences and then use a scoring function to find important words to manipulate. Through their experiments under both settings, they show that an adversary can deceive multiple real-world online systems with the generated adversarial texts.

\section{Methodology}

\subsection{Threat Model}
Hate speech detection is being used in the security landscape in an increasingly wider range of applications. Consequently, understanding the security properties of the mechanisms that are deployed for hate speech detection has become crucial. The extent to which we can craft adversarial samples influences the applications of hate speech defence models.
 We assume in this paper that the adversary has black-box access to the hate speech detection model. The adversary is assumed to be operating under the following constraints:
 \begin{itemize}
     \item The adversary has only query access to the model. Specifically, the adversary can only query the hate speech detection model API with a sample and will get a score in response (3 scores in case of Hate Sonar). This score is on a scale of 0 to 1 where 0 denotes not hateful and 1 denotes most hateful. Perspective API \cite{perspective_api_website} can be accessed over HTTPS protocol while the HateSonar \cite{hatesonar_github} API is exposed as a python library distributed through PyPI \cite{pypi_website}. More details about the API contracts in the Experimental Setup section.
     \item The adversary has no knowledge of the architecture of the hate speech detection model.
     \item The adversary has no knowledge of the dataset used to train the model.
     \item The adversary has rate-limited access to the Perspective API endpoint. We were able to access 50 Query Per Second rate-limit for the Perspective API endpoint without many efforts.
 \end{itemize}
 
In essence, the adversary can only query the model with a sample and get back the hateful/ toxicity score. It has no other knowledge of the model. Needless to say, the adversary has no access to any gradients of the hate speech detection models. Our attack surface would be online social media platforms since these are the primary targets for attackers and often employ hate speech detection models for curbing hate speech.

\subsection{Dataset description and analysis}
We used the hate speech dataset by Mondal et al \cite{mondal_dataset}. This dataset contains total 20,705 posts from Twitter collected in 2014-2015. The original dataset contains three columns:
\begin{itemize}
    \item \textbf{Tweet Id}: The unique id of the tweet assigned by Twitter.
    \item \textbf{Hate targets extracted from the tweet text}: Contains the groups of people who are the target of that particular tweet.
    \item \textbf{Hate categories}: Manually labelled hate categories. Table \ref{tab:hate_categories}.
\end{itemize}
However, upon request to the authors of \cite{mondal_dataset}, we obtained the tweet texts corresponding to the Tweet Ids in the dataset. Throughout our work, we mostly work on these tweet texts and ignore the other information in the dataset.

\subsubsection{Dataset analysis on Perspective API} \label{dataset_analysis_for_perspective_api}
\begin{figure}
    \centering
    \includegraphics[scale=0.2]{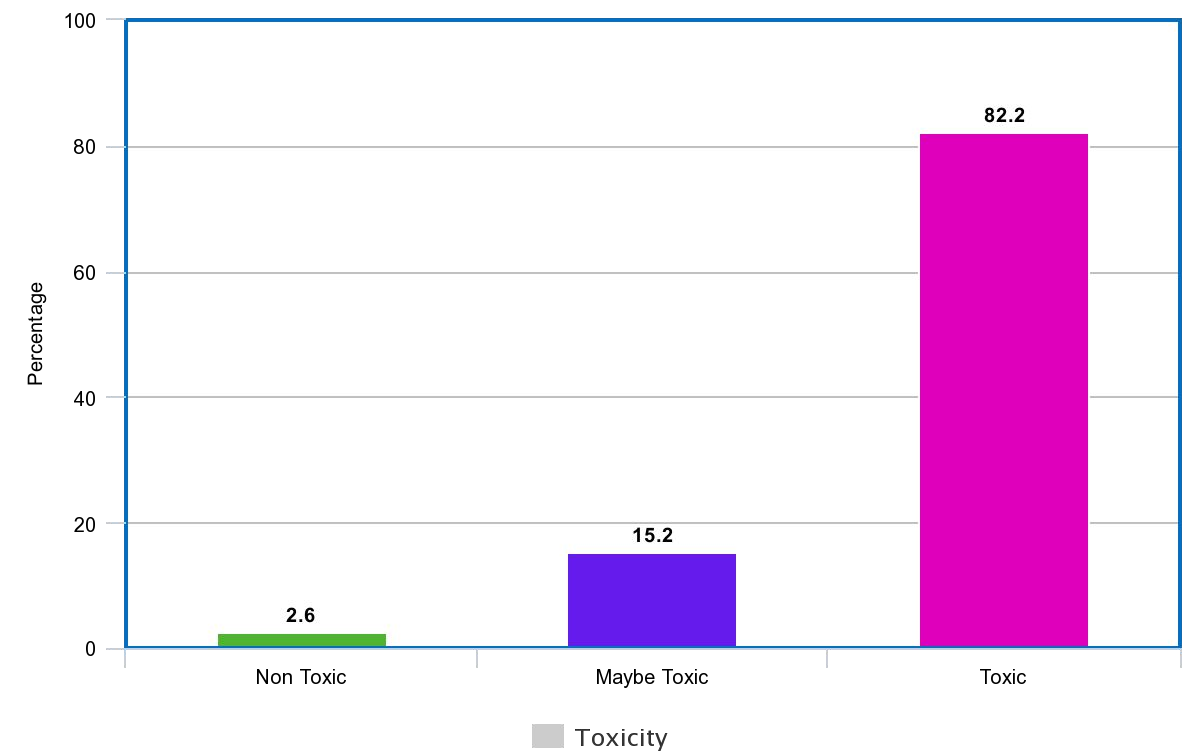}
    \caption{Category distribution of dataset according to Perspective API}
    \label{fig:papi_dataset_analysis_bar_plot}
\end{figure}

\begin{figure}
    \centering
    \includegraphics[scale=0.13]{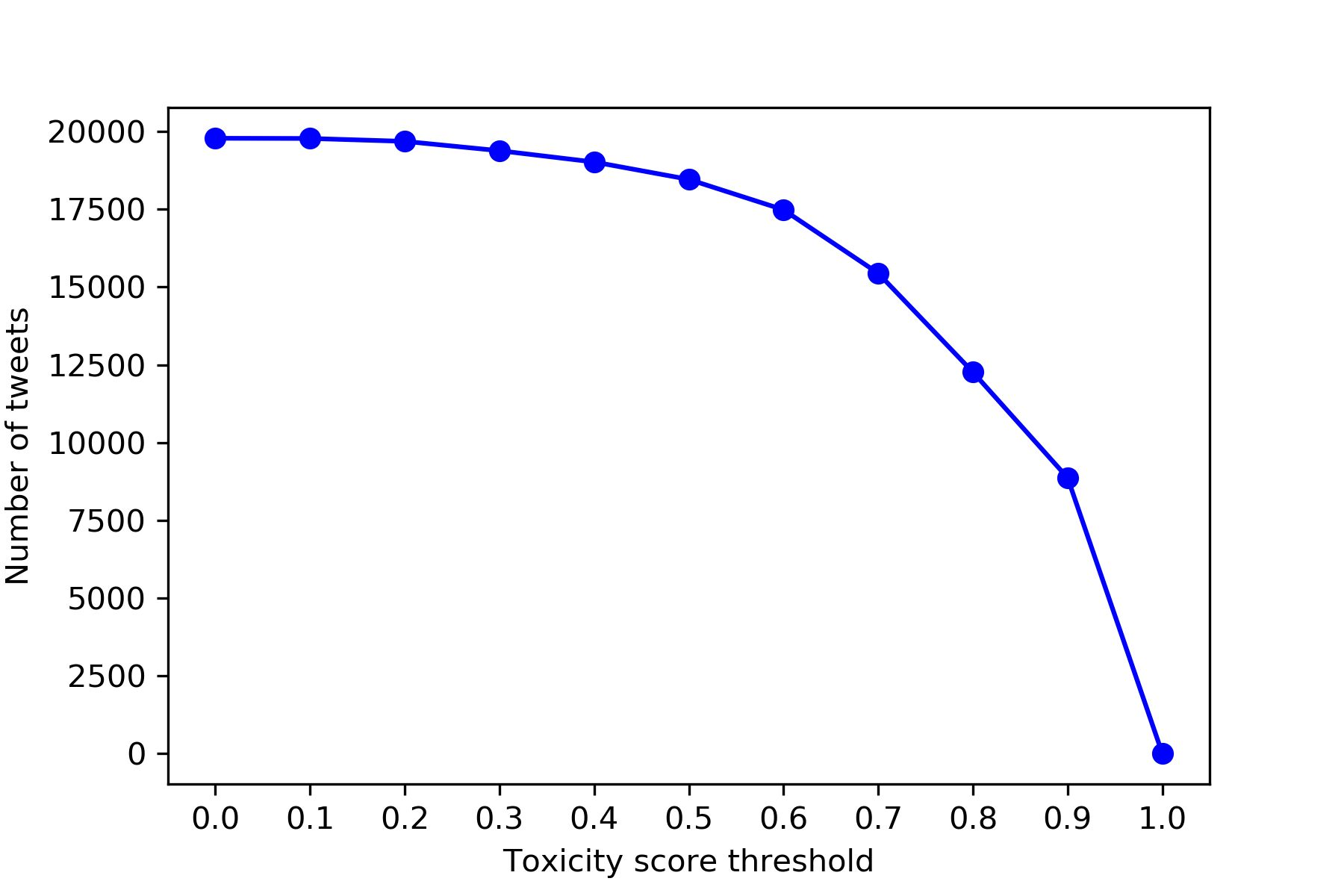}
    \caption{How the toxicity of the dataset varies with toxicity threshold for Perspective API.}
    \label{fig:papi_dataset_analysis_threshold_distribution}
\end{figure}

We obtained the toxicity for each tweet in the dataset by querying Perspective API. Further, we thresholded the toxicity values using the thresholds mentioned in Section \ref{details_of_papi_and_hatesonar}. Figure \ref{fig:papi_dataset_analysis_bar_plot} shows the category distribution. Further \ref{fig:papi_dataset_analysis_threshold_distribution} shows how the toxicity of the dataset varies with the toxicity threshold for Perspective API. From these two figures, we can observe that most of the tweets in the dataset are toxic according to Perspective API.

{}

{}

\subsubsection{Dataset analysis on HateSonar}
\begin{figure}
    \centering
    \includegraphics[scale=0.125]{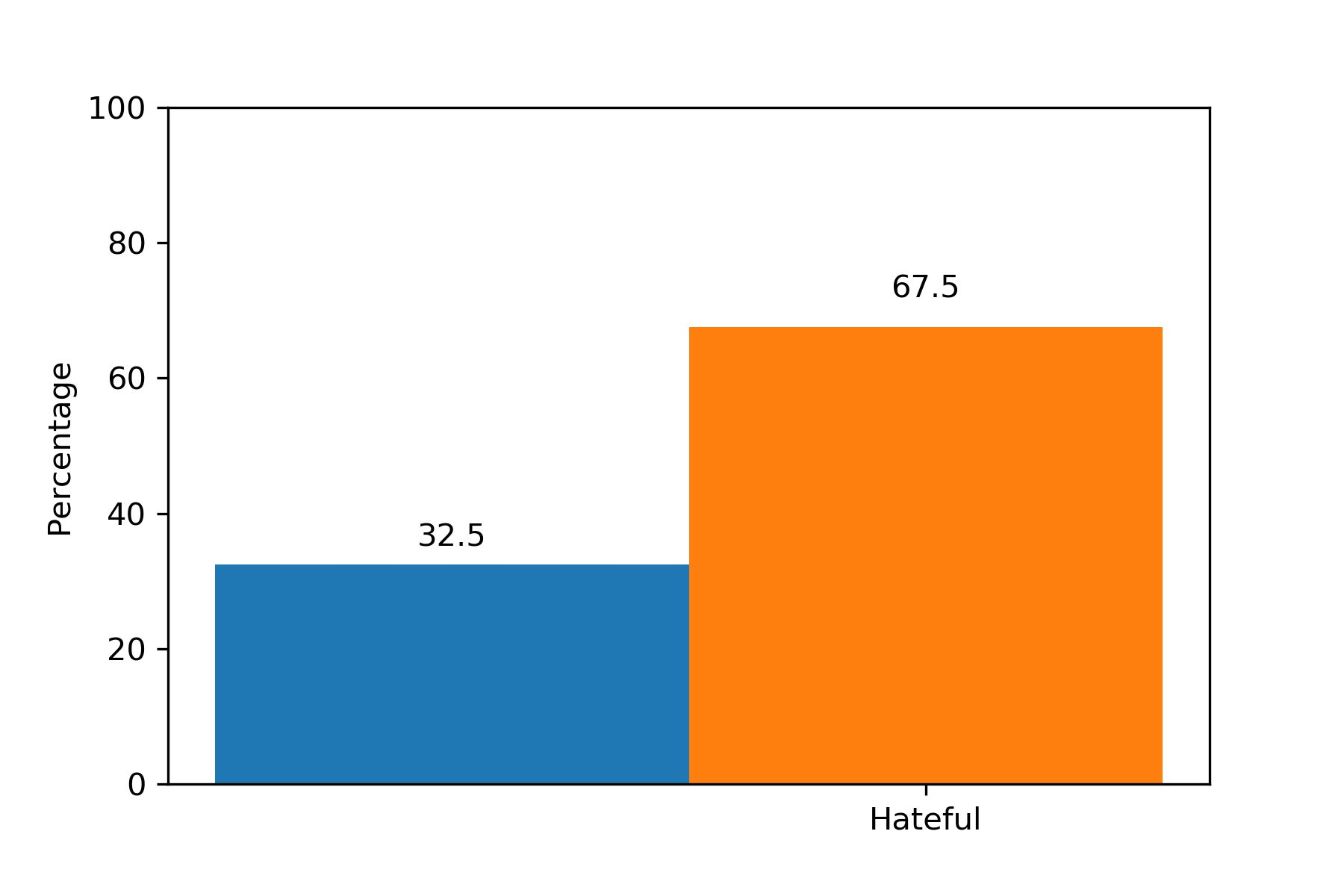}
    \caption{Category distribution of dataset according to HateSonar}
    \label{fig:hatesonar_dataset_analysis_bar_plot}
\end{figure}

Similar to Section \ref{dataset_analysis_for_perspective_api}, the category for each example in the dataset was found by querying the HateSonar model and by using the categorization methodology mentioned in Section \ref{details_of_papi_and_hatesonar}. Figure \ref{fig:hatesonar_dataset_analysis_bar_plot} shows the result.

{}

\begin{table}
  \caption{Hate categories with example of hate targets. \cite{mondal_dataset}}
  \label{tab:hate_categories}
  \begin{tabular}{ccl}
    \toprule
    Categories & Examples of hate targets\\
    \midrule
    Race & nigga, nigger, black people, white people \\
    Behavior & insecure people, slow people, sensitive people \\
    Physical & obese people , short people, beautiful people \\
    Sexual orientation & gay people, straight people \\
    Class & ghetto people, rich people \\
    Gender & pregnant people, cunt, sexist people \\
    Ethnicity & chinese people, indian people, paki \\
    Disability & retard, bipolar people \\
    Religion & religious people, jewish people \\
    Other & drunk people, shallow people \\
  \bottomrule
\end{tabular}
\end{table}

\subsection{Experimental setup}
\subsubsection{Details about Perspective API and HateSonar} \label{details_of_papi_and_hatesonar}
\textbf{Perspective API} \cite{perspective_api_website} is an online service owned by Google Inc. Behind this service is a deep learning model based n the CNN architecture. It uses Glove word vector embedding and is trained on Wikipedia'2014 and Gigaword 5 datasets. These datasets contains 6 billion tokens and 300K vocab. The data set includes over 100k labeled discussion comments from English Wikipedia. Each comment was labeled by multiple annotators via Crowdflower on whether it is a toxic or healthy contribution \cite{perspective_background}. We requested developer access to this service to be able to use it's HTTP API. Initially, we were granted developer API access with a rate-limit of 10 queries per second (QPS). However, upon request to the Perspective API team, this was later increased to 50 QPS. As per the API contract of Perspective API, we can pass in a text string within 3000 bytes to the API using an HTTP POST request and the response will contain the overall toxicity score of the text string that was passed in the input request. Further, the HTTP API also supports a \textit{span annotation} feature. This feature returns a sentence level toxicity of the input text. For example, if the input sentence is:

\textit{'The quick brown fox jumped over the fence. There are many sheep in the farm'}

then, in the response, the API, along with an overall toxicity score, will return two sentence-level toxicity scores for each of the two sentences in the above example.
During our experiments, we also observed that the API did not return any toxicity score for certain inputs. More details regarding this in the \textit{Error handling} section.

For our analysis, we thresholded the toxicity score returned by Perspective API into three buckets.
\begin{itemize}
    \item \textbf{Non-toxic}: 0.00 to 0.33
    \item \textbf{Maybe-toxic}: 0.33 to 0.66
    \item \textbf{Toxic}: 0.66 to 1.00
\end{itemize}

\textbf{HateSonar} \cite{hatesonar_github} is an open-source Python library. This model was trained on the dataset mentioned in \cite{davidson_dataset}. This library hosts a model in itself, i.e. it does not make any HTTP call over the network for making deductions. Hence, there are no rate-limits for querying this model. The authors note that although it might be possible to get the gradients or have white-box access to this model through the library, this information was not used for crafting adversarial samples in this work. The implementation of this model uses Logistic regression with l2 regularization. The overall precision, recall and F1 score for this model are 0.91, 0.90 and 0.90 as mentioned in \cite{davidson_dataset}.

Similar to Perspective API, HateSonar returns scores for a given input. However, the response of HateSonar differs from Perspective API in the sense that it returns the confidence scores for three classes, i.e. \textit{hate\_speech}, \textit{offensive\_language} and \textit{neither}. For the purpose of our evaluation, we assume the text to be hateful if the confidence of \textit{neither} is not the highest among the three classes. Although this assumption makes it harder for our perturbation to perform better, it makes the evaluation fairer. One of the intentions behind doing this was to align the output of HateSonar with that of Perspective API.

To explain better, for Hate Sonar responses, we thresholded the response as follows:
\begin{itemize}
    \item \textbf{Non-toxic}, if the class \textit{neither} has the highest confidence score out of the three classes.
    \item \textbf{Toxic}, if the class \textit{neither} does not have the highest confidence score out of the three classes.
\end{itemize}

\subsubsection{Finding the most toxic word in the example}
We tried two approaches for determining the most toxic word in the tweet. In the first approach, we leveraged the \textit{span annotation} feature of Perspective API. To achieve this, we added a period before every space character in the tweet and capitalized every alphabetical character immediately after space. The intention behind doing this was to make Perspective API believe that every word in the tweet is a separate sentence thereby fooling it into returning the toxicity score of every word. For example, a sentence like:

\textit{The quick brown fox jumped over the fence.}

was changed to:

\textit{The. Quick. Brown. Fox. Jumped. Over. The. Fence.}

However, upon manually inspecting the results we observed that what appeared to be the most toxic word often did not have the highest toxicity scores. One possible explanation for this behaviour is that the  Perspective API might be using the context of the sentence for determining toxicity scores. In other words, since the \textit{span annotation} feature looks at each sentence ('word' in our case) in isolation, it was not able to correctly ascribe a toxicity score.

\begin{figure}[h]
  \centering
  \includegraphics[width=\linewidth]{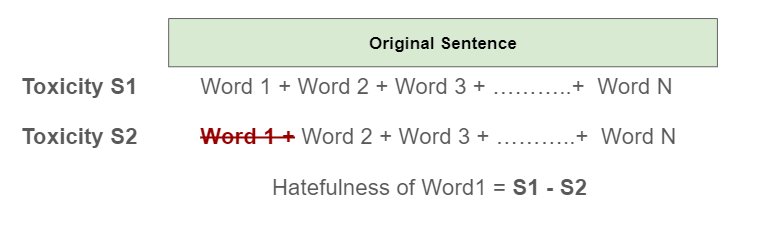}
  \caption{Edit distance evaluations for perturbations on Perspective API and Hate Sonar}
 \label{fig:candidate_diagram}
\end{figure}

Hence, we changed our approach to the following as also described by Figure \ref{fig:candidate_diagram}:
\begin{enumerate}
    \item Get the toxicity score of the original tweet by querying Perspective API.
    \item Tokenize the tweet into words.
    \item For each word:
    \begin{enumerate}
        \item Remove it from the original tweet.
        \item Get the toxicity score of this 'word-removed-tweet' by querying Perspective API.
        \item Assign the toxicity of the removed word as the difference in the toxicities of the original tweet and the 'word-removed-tweet'.
    \end{enumerate}
\end{enumerate}

Upon manual inspection of the results of this approach, we observed that the word-level toxicities were in alignment with our perception of the toxicity of words.

The authors note that the second approach mentioned above did not work with HateSonar since the word-level toxicities computed using the HateSonar API did not align with our perception of the toxicity of words. Therefore, to select a candidate word for perturbation for HateSonar evaluations, the word-level toxicities computes using Perspective API were used.

\subsubsection{Description of perturbations}
\label{description_of_perturations}

\begin{figure}
    \centering
    \includegraphics[scale=0.15]{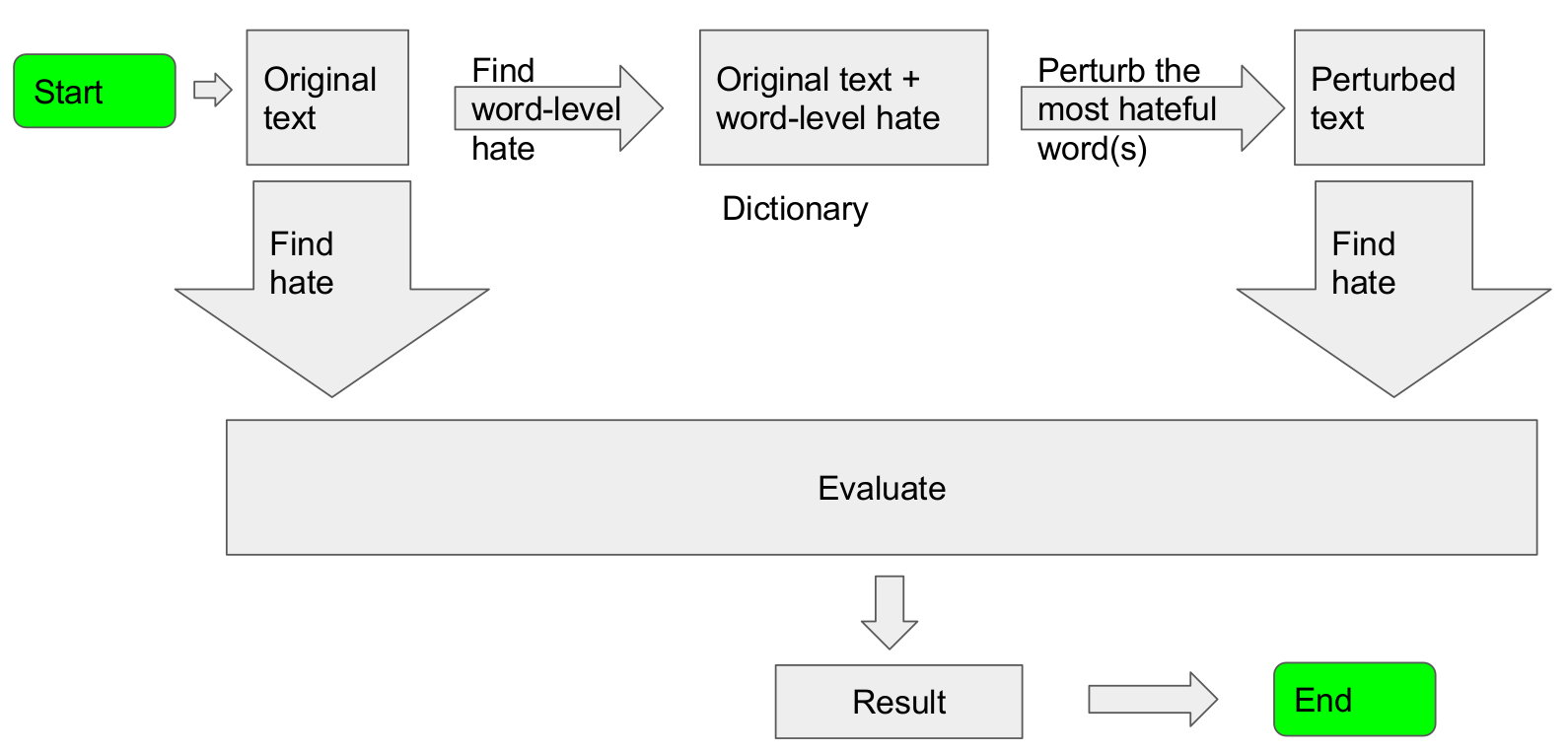}
    \caption{Process diagram for our approach.}
    \label{fig:process_diagram}
\end{figure}

The toxicities for all the tweets in the dataset were computed by querying each of them with Perspective API (or Hate Sonar). Further, after perturbing each tweet using one of the perturbations approaches described below, the toxicity for each perturbed tweet was computed again by querying with Perspective API (or HateSonar). See Figure \ref{fig:process_diagram}.

{}

\begin{itemize}
    \item \textbf{Leet speak}: Leet speak is a system of modified spellings used primarily on the internet \cite{leetspeak_definition}. For example, the word \textit{noob} would be represented in leet speak as \textit{n00b}. On similar lines, we apply leet speak to the most toxic word(s) in the sentence. To apply leet speak to a word, we have defined a mapping from normal English alphabetical characters (i.e. a-z and A-Z) to a list of unicode characters. For example the alphabet \textit{a} will be replaced by the \textit{Cyrillic small letter A}, and so on. The entire mapping is described in Table \ref{tab:leet_speak_mapping}. For example, see Figure \ref{fig:leet_screenshot}.
    \begin{figure}[h]
        \centering
        \includegraphics[width=\linewidth]{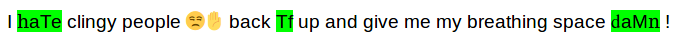}
        \caption{Example of a sentence perturbed by leet speak. Highlighted words have been perturbed.}
        \label{fig:leet_screenshot}
    \end{figure}{}
    
    \item \textbf{Insertion of typos}: In this perturbation, we introduce typos (i.e. spelling mistakes) in the original text. The two most hateful word(s) are targeted for introducing typos. Specifically, there are three possible scenarios that need to be handled for introducing a typo in the word:
    \begin{itemize}
        \item \textbf{Length of the word is less than 4 characters: }Do not perturb the word.
        \item \textbf{Length of the word is an even number: }Interchange the middle two characters in the word. For example \textit{hate} would become \textit{htae}.
        \item \textbf{Length of the word is an odd number: }Interchange the two characters surrounding the middle character. For example \textit{walks} would become \textit{wklas}.
    \end{itemize}{}
    
    \item \textbf{Insertion of underscores}: In this perturbation, every white space character in the sentence would get replaced by an underscore character.
    \\
    For example:
    \\
    \textit{The quick brown fox jumped over the fence.}
    \\
    would get changed to:
    \\
    \textit{The\_quick\_brown\_fox\_jumped\_over\_the\_fence.}
    
    \item \textbf{Removal of whitespace}: In this perturbation, every white space character in the sentence would be removed.
    \\
    For example:
    \\
    \textit{The quick brown fox jumped over the fence.}
    \\
    would get changed to:
    \\
    \textit{Thequickbrownfoxjumpedoverthefence.}
    
    \item \textbf{Insertion of zero width whitespace}: In this perturbation, we add the \textit{zero width white space} Unicode character. The Unicode value of this character is U+200B. This \textit{zero width white space} character was inserted 5 times between each character of the most toxic word in the sentence. Visually, the original and perturbed text look identical leading to no change in readability for this perturbation.
    
    \item \textbf{Composite attack 1 (Insertion of underscores + Leetspeak) :} In this attack, we apply two types of perturbations simultaneously to a single input text, i.e. insertion of underscores and Leetspeak.
    
    \item \textbf{Composite attack 2 (Zero width white space + Leetspeak) :} Similar to \textit{Composite attack 1}, we apply two types of perturbations simultaneously to a single input text, i.e. insertion of zero width white space and Leetspeak.
\end{itemize}

\subsubsection{Error handling} \label{error_handling}
For some perturbed texts, Perspective API was unable to return any toxicity score. Specifically, the response from Perspective API said: \textit{'
Sorry! Perspective needs more training data to work in this language'}. The authors observed that this happened for sentences which had a higher amount of perturbation. For instance, Perspective API exhibited this behaviour for sentences perturbed heavily using Leet Speak. This might be happening because our implementation of Leet Speak uses quite a few of Unicode characters which look similar to English alphabets.

\subsection{Evaluation Metrics}

\subsubsection{Metrics to measure the effectiveness of perturbations}
\begin{itemize}
    \item \textbf{Mean change in toxicity:} This metric measures how much the mean toxicity of the dataset was changed because of a perturbation and is only applicable to Perspective API. In other words, the toxicity of the entire dataset is initially calculated using Perspective API. A mean of all these toxicities is then calculated. A similar process is done for the perturbed dataset to get a mean toxicity value for the perturbed dataset. The difference in these two computed mean values is termed as the mean change in toxicity.
    \item \textbf{Category shift score:} As mentioned in section \ref{details_of_papi_and_hatesonar}, the category of hatefulness is computed for a given sample by querying Perspective API (or HateSonar), i.e. \textit{Toxic}, \textit{Maybe Toxic} or \textit{Non Toxic}. The \textit{category shift score} is defined as the the percentage of the total examples in the dataset that went from the \textit{Toxic} category to any other category. A similar definition would hold true for HateSonar.
    \item \textbf{Modified category shift score:} This metric is only applicable for Perspective API since it is possible that Perspective API sometimes would not return the toxicity value (See section \ref{error_handling}) for a given input text. Thus, \textit{modified shift score} is defined as the percent of total examples in the dataset that went from the \textit{Toxic} category to any other category or for whom Perspective API did not return a toxicity score. In other words, this metric is the sum of the \textit{category shift score} and the percent of samples not recognized by Perspective API.
\end{itemize}{}

\subsubsection{Metrics to measure the amount of perturbations}
\begin{itemize}
\item \textbf
{Edit Distance:}
Edit distance is a way of quantifying how
dissimilar two strings (e.g., sentences) are by counting the minimum number of operations required to transform one string to the other. Specifically, different definitions of the edit distance use different sets of string operations. In our experiment, we use the most common metrics, i.e., the Levenshtein distance\cite{EditDistance}, whose operations include removal, insertion, and substitution of characters in the string.
\item \textbf{Human Evaluation:}
While an extensive user study to measure the semantic similarity between the original and perturbed texts is not conducted in this work, we rely on peer evaluation while presenting the findings in the class.
\end{itemize}

\section{Results}
\begin{figure}[h]
  \centering
  \includegraphics[width=\linewidth]{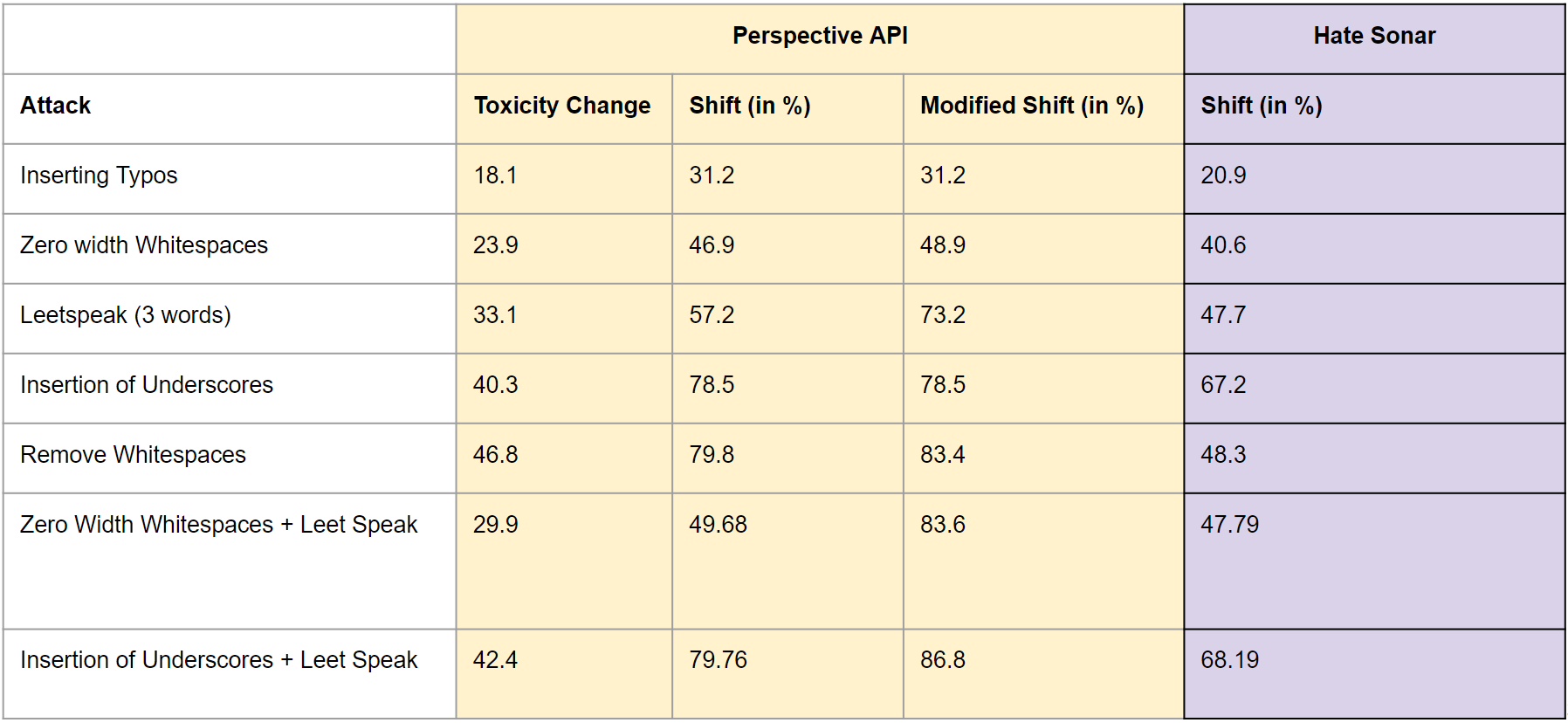}
  \caption{Evaluations for perturbations on Perspective API and Hate Sonar}
  \label{fig:evaluations_diagram}
\end{figure}

Figure \ref{fig:evaluations_diagram} illustrates the performance of various perturbations on our evaluation metrics as described before. Among homogeneous attacks, while the insertion of typos achieves the worst performance, insertion of underscores and removal of white spaces achieves the best results. The better performance for white space manipulation attacks might give us some insight about the tokenization process of the models being attacked. One of the reasons that the models failed can be because they considered the whole string as a single word. Among the composite attacks, Insertion of Underscores + Leetspeak resulted in the best performance. The resulting shifts and change were higher than both the insertion of underscores and Leetspeak attacks considered separately. The figures for the individual perturbation results are in the appendix. See section \ref{appendix}.

Figure \ref{fig:edit_distance_diagram} illustrates the edit distance evaluations for all the perturbations. Since the insertion of typos perturbation involves just swaps of some characters, it results in the minimum edit distance between the original and perturbed sentences. In our experiments, we concluded that inserting a single zero width white space does not suffice the aim to reduce the hate content. So, we added multiple zero width white spaces before the target word. This has led to extremely high edit distance values.

Further, we displayed different sentences perturbed with all kinds of attacks to our peers in the class during the final project presentation. It was the unanimous opinion of the class that even after the perturbations, all the displayed sentences had retained their hateful meaning completely.

\begin{figure}[h]
  \centering
  \includegraphics[width=\linewidth]{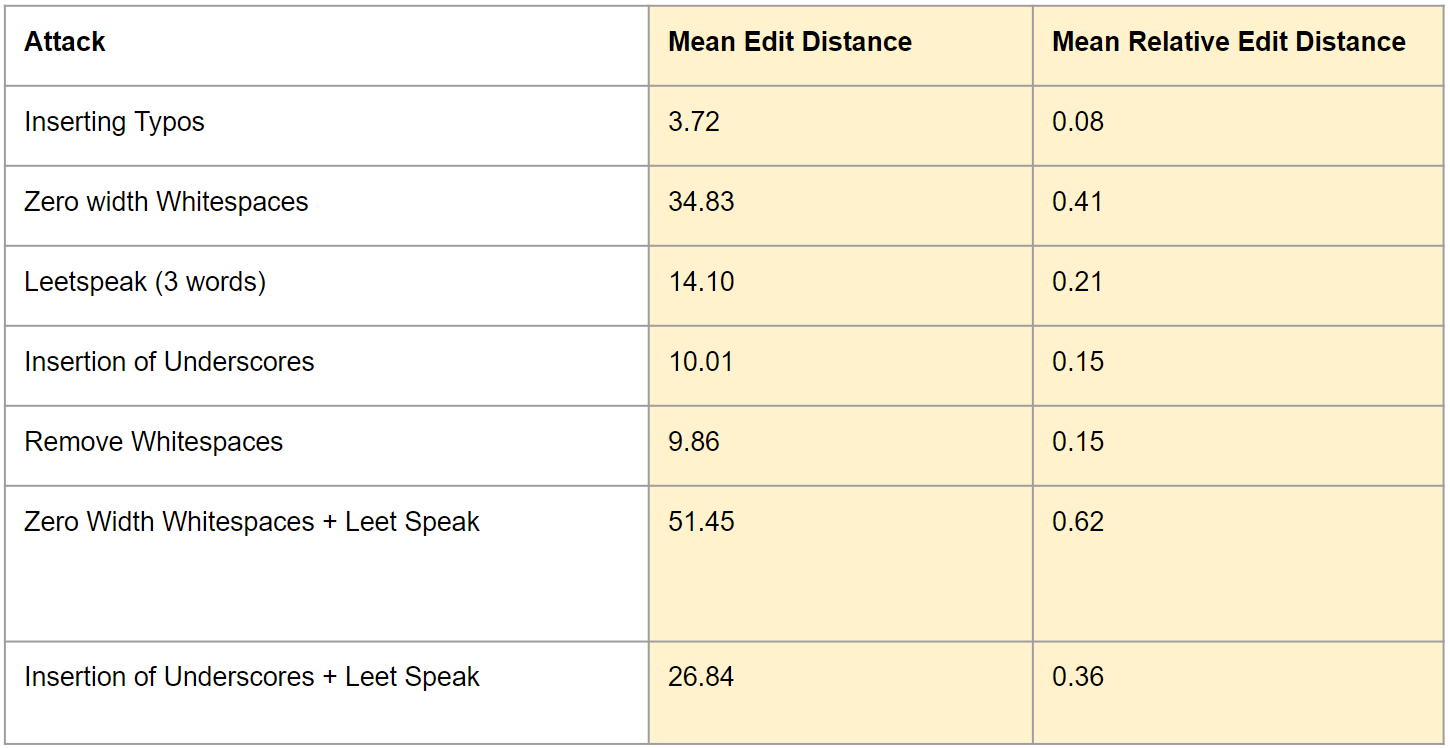}
  \caption{Edit distance evaluations for perturbations on Perspective API and Hate Sonar}
 \label{fig:edit_distance_diagram}
\end{figure}

\section{Proposed defenses}

\subsection{Leet speak}
    We use a mapping from English characters to Unicode character (e.g. Cyrillic alphabet, Greek alphabet, Latin alphabet, etc). Thus, if there exists an inverse dictionary to map from these Unicode characters to the regular English alphabet, then the during prediction time, the input string can be sanitized to replace all Unicode alphabets available in the inverse dictionary with the regular English alphabet. Constructing such an inverse dictionary should be trivial since we already have the original dictionary.

\subsection{Insertion of typos}
An auto-correct software can be used to sanitize the input string before making the prediction. Although this might miss some cases, most of the hateful content should be detected using this method.

\subsection{Insertion of underscores}
Inserting underscores significantly degrades the performance of both Perspective API and HateSonar. Intuitively, this might be because both these models must be using white space-based tokenization for tokenizing sentences into words. Therefore, updating this tokenization logic to tokenize on both white space and underscores should help significantly reduce the impact of this attack. The authors note that in case there are any intentional underscores in the original text, this updated tokenization logic would wrongly tokenize on it.

\subsection{Zero width white space}
Removing all zero width white space characters using regex matching is proposed to be a good defence against this attack.

\subsection{Removal of white space}
The famous word-break algorithm can be used to defend against this attack. In short, the word break algorithm can be described as:
\textit{Given a String and a dictionary of words, write a program that returns true if the given string can be formed by concatenating one or more of the words in the dictionary.} The time complexity of this algorithm is O(m x s) where m is the number of characters in the perturbed string which needs to be word-broken. And s in the number of characters in the longest word in the provided dictionary.
The authors note that by using this algorithm, multiple possible reconstructions of original sentences can be formed given a perturbed sentence.

\subsection{Composite attacks}
Respective combination of defences can be employed against the two composite attacks mentioned in Section \ref{description_of_perturations}.

\section{Limitations and Future Work}
\begin{itemize}
    \item \textbf{White Box attacks:}
    
    During this work, we have only focused on black box based attacks. White-box attacks find or approximate the worst-case attack for a particular model and input based on the Kerckhoff’s principle\cite{TextBugger}. Therefore, white-box attacks can expose a model’s worst case vulnerabilities. Thus, in the uture we would like to extend the work to white box setting.
    
    \item \textbf{Use of other data sets:}
    
    We have based all our evaluations on the \cite{mondal_dataset}. To establish the generalisability of our perturbation based attacks, we would like to extend the work to encompass more data sets.
    
    \item \textbf{API Rate Limits}
    
    As we discussed in the introduction, most of the deep learning models accessible through APIs have a rate limit associated with them. This limitation causes an issue for large datasets and large texts.
    
    \item \textbf{Dependence of Hate Sonar on Perspective API}
    
  The Hate Sonar API returns a classification between Hate, Offence and Neither. Since our approach is based on finding the most toxic words(s), we use the dictionary created using the Perspective API to find the candidate words. But, we still feel that even with this limitation, the design gives us a fair idea of the performance of various perturbations across models.

\end{itemize}

\section{Conclusion}
We came up with 3 classes of perturbations totalling 7 attacks. For homogeneous attacks, insertion of underscores and removal of white spaces performed the best while the combination of insertion of underscore and leet speak performed the best across all categories.

\bibliographystyle{ACM-Reference-Format}
\bibliography{bibliography}

\section{appendix} \label{appendix}

\begin{figure}[H]
    \centering
    \includegraphics[width=\linewidth]{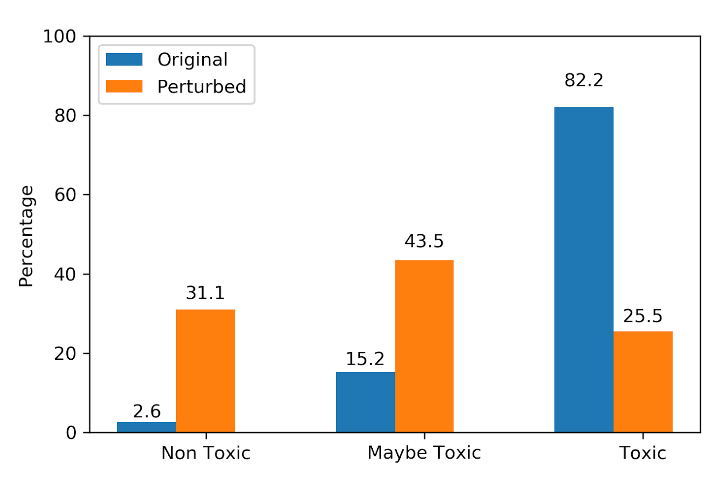}
    \caption{Original and resulting toxicities for Leet speak perturbation for Perspective API}
    \label{fig:leet_papi}
\end{figure}{}

\begin{figure}[H]
    \centering
    \includegraphics[width=\linewidth]{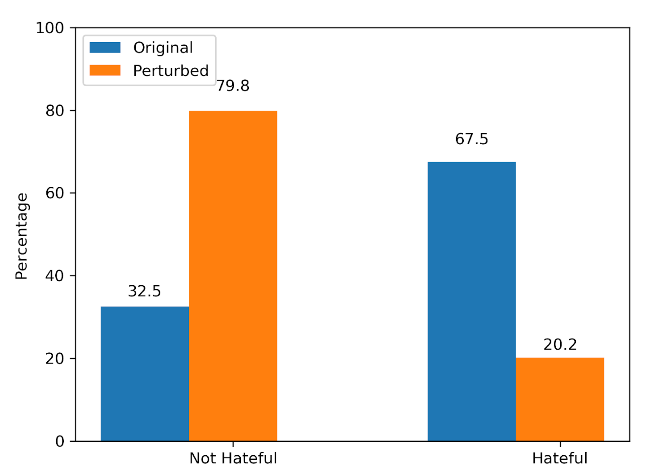}
    \caption{Original and resulting toxicities for Leet speak perturbation for HateSonar}
    \label{fig:leet_hatesonar}
\end{figure}{}

\begin{figure}[H]
    \centering
    \includegraphics[width=\linewidth]{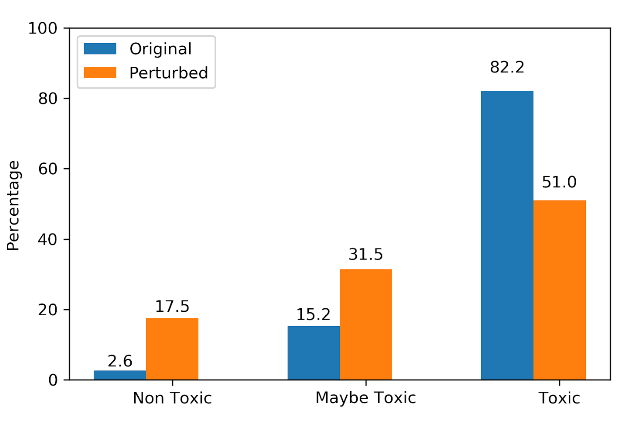}
    \caption{Original and resulting toxicities for Typo perturbation for Perspective API}
    \label{fig:typo_papi}
\end{figure}{}

\begin{figure}[H]
    \centering
    \includegraphics[width=\linewidth]{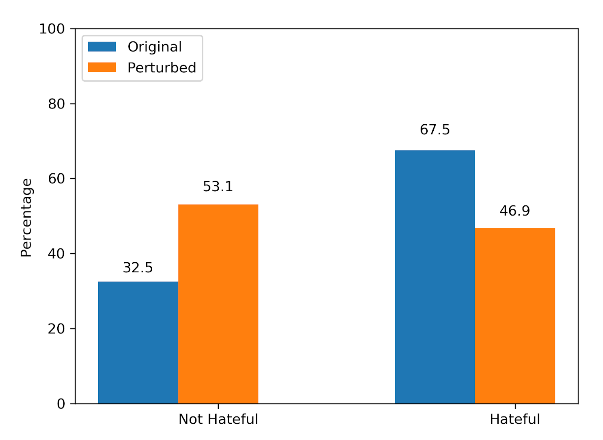}
    \caption{Original and resulting toxicities for Typo perturbation for HateSonar}
    \label{fig:typo_hatesonar}
\end{figure}{}

\begin{figure}[H]
    \centering
    \includegraphics[width=\linewidth]{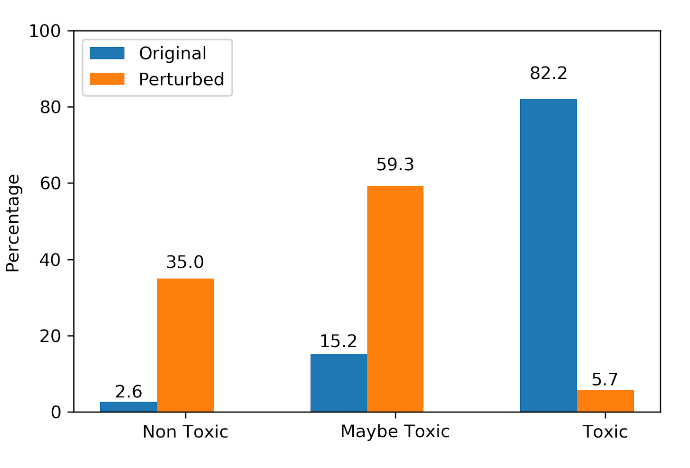}
    \caption{Original and resulting toxicities for underscore perturbation for Perspective API}
    \label{fig:underscore_papi}
\end{figure}{}

\begin{figure}[H]
    \centering
    \includegraphics[width=\linewidth]{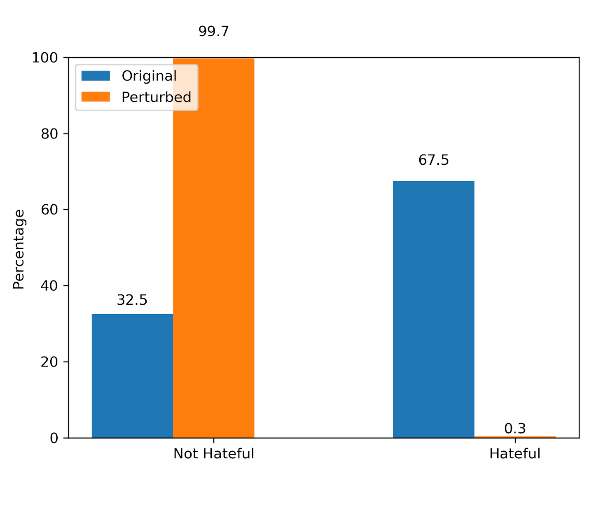}
    \caption{Original and resulting toxicities for underscore perturbation for HateSonar}
    \label{fig:underscore_hatesonar}
\end{figure}{}

\begin{figure}[H]
    \centering
    \includegraphics[width=\linewidth]{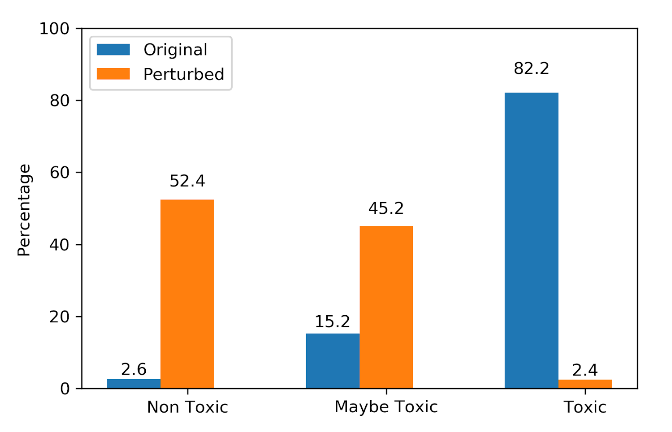}
    \caption{Original and resulting toxicities for removal of white space perturbation for Perspective API}
    \label{fig:removal_papi}
\end{figure}{}

\begin{figure}[H]
    \centering
    \includegraphics[width=\linewidth]{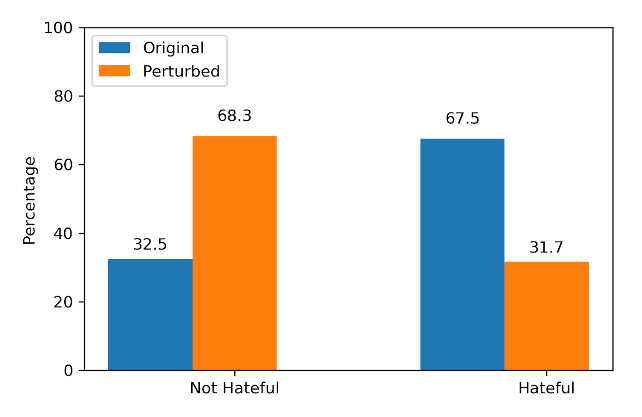}
    \caption{Original and resulting toxicities for removal of white space perturbation for HateSonar}
    \label{fig:removal_hatesonar}
\end{figure}{}

\begin{figure}[H]
    \centering
    \includegraphics[width=\linewidth]{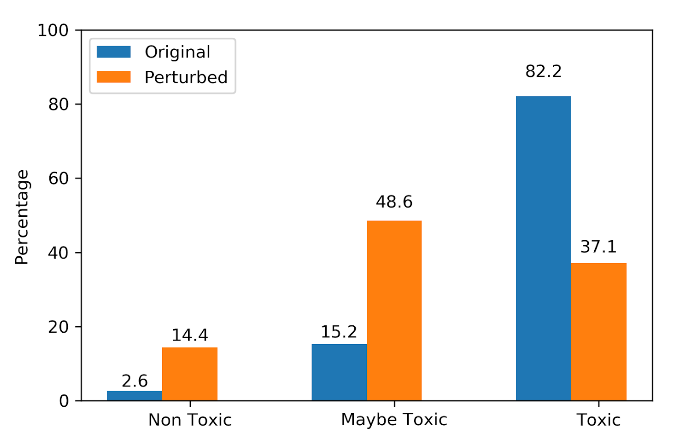}
    \caption{Original and resulting toxicities for zero width white space perturbation for Perspective API}
    \label{fig:zwsp_papi}
\end{figure}{}

\begin{figure}[H]
    \centering
    \includegraphics[width=\linewidth]{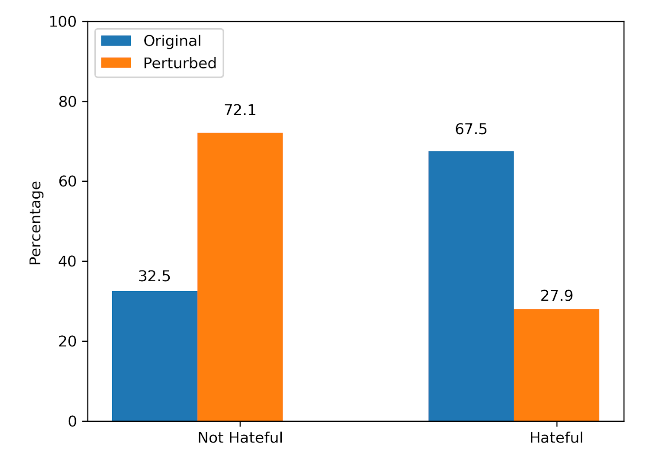}
    \caption{Original and resulting toxicities for zero width white space perturbation for HateSonar}
    \label{fig:zwsp_hatesonar}
\end{figure}{}

\begin{figure}[H]
    \centering
    \includegraphics[width=\linewidth]{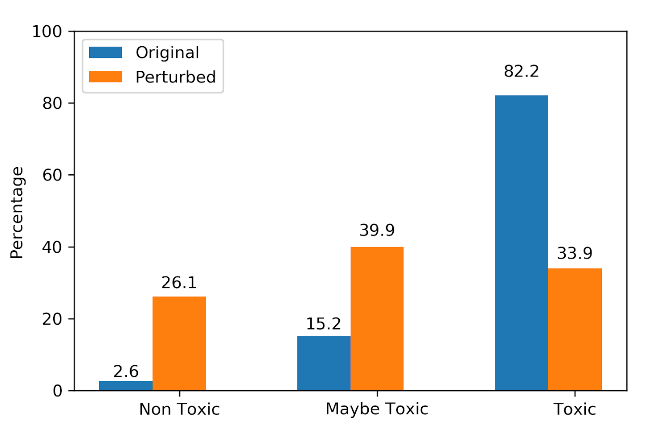}
    \caption{Original and resulting toxicities for composite (zero width white space + leet speak) perturbation for Perspective API}
    \label{fig:zwsp_leet_papi}
\end{figure}{}

\begin{figure}[H]
    \centering
    \includegraphics[width=\linewidth]{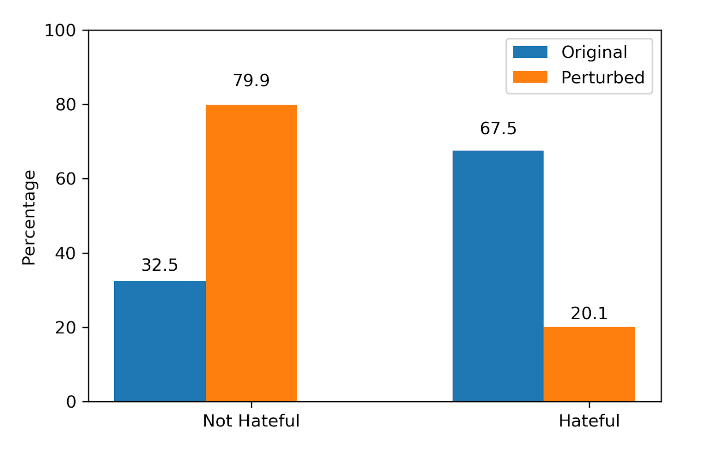}
    \caption{Original and resulting toxicities for composite (zero width white space + leet speak) perturbation for HateSonar}
    \label{fig:zwsp_leet_hatesonar}
\end{figure}{}

\begin{figure}[H]
    \centering
    \includegraphics[width=\linewidth]{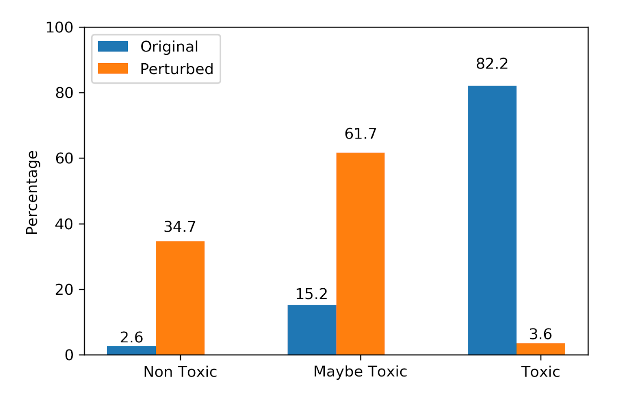}
    \caption{Original and resulting toxicities for composite (underscore + leet speak) perturbation for Perspective API}
    \label{fig:underscore_leet_papi}
\end{figure}{}

\begin{figure}[H]
    \centering
    \includegraphics[width=\linewidth]{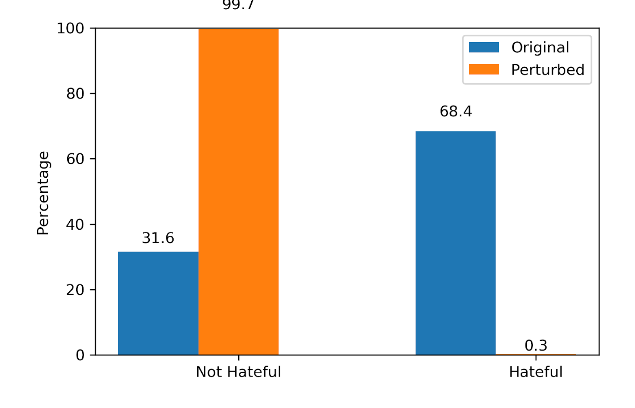}
    \caption{Original and resulting toxicities for composite (underscore + leet speak) perturbation for HateSonar}
    \label{fig:underscore_leet_hatesonar}
\end{figure}{}

\begin{table}[h]
  \caption{Character mapping for leet speak.}
  \label{tab:leet_speak_mapping}
  \begin{tabular}{p{1cm} p{7cm}}
    \toprule
    Target & Replacement character name\\
    \midrule
    'a' & 'CYRILLIC SMALL LETTER A' \\
    'A' & 'CYRILLIC CAPITAL LETTER A' \\
    'b' & 'CYRILLIC CAPITAL LETTER SOFT SIGN' \\
    'B' & 'CYRILLIC CAPITAL LETTER VE' \\
    'c' & 'CYRILLIC SMALL LETTER ES' \\
    'C' & 'CYRILLIC CAPITAL LETTER ES' \\
    'd' & 'CYRILLIC SMALL LETTER KOMI DE' \\
    'D' & 'CHEROKEE LETTER A' \\
    'e' & 'CYRILLIC SMALL LETTER IE' \\
    'E' & 'CYRILLIC CAPITAL LETTER IE' \\
    'f' & 'LATIN SMALL LETTER LONG S WITH HIGH STROKE' \\
    'F' & 'LISU LETTER TSA' \\
    'g' & 'ARMENIAN SMALL LETTER CO' \\
    'G' & 'CYRILLIC CAPITAL LETTER KOMI SJE' \\
    'h' & 'CYRILLIC SMALL LETTER SHHA' \\
    'H' & 'CYRILLIC CAPITAL LETTER EN' \\
    'i' & 'CYRILLIC SMALL LETTER BYELORUSSIAN-UKRAINIAN I' \\
    'I' & 'CYRILLIC SMALL LETTER BYELORUSSIAN-UKRAINIAN I' \\
    'j' & 'CYRILLIC SMALL LETTER JE' \\
    'J' & 'CYRILLIC CAPITAL LETTER JE' \\
    'k' & 'CYRILLIC CAPITAL LETTER KA' \\
    'K' & 'CYRILLIC CAPITAL LETTER KA' \\
    'l' & 'CHEROKEE LETTER TLE' \\
    'L' & 'CHEROKEE LETTER TLE' \\
    'm' & 'CYRILLIC CAPITAL LETTER EM' \\
    'M' & 'CYRILLIC CAPITAL LETTER EM' \\
    'n' & 'ARMENIAN SMALL LETTER VO' \\
    'N' & 'GREEK CAPITAL LETTER NU' \\
    'o' & 'CYRILLIC SMALL LETTER O' \\
    'O' & 'CYRILLIC CAPITAL LETTER O' \\
    'p' & 'CYRILLIC SMALL LETTER ER' \\
    'P' & 'CYRILLIC CAPITAL LETTER ER' \\
    'q' & 'CYRILLIC SMALL LETTER QA' \\
    'Q' & 'TIFINAGH LETTER YARR' \\
    'r' & 'CYRILLIC SMALL LETTER GHE' \\
    'R' & 'LISU LETTER ZHA' \\
    's' & 'CYRILLIC SMALL LETTER DZE' \\
    'S' & 'CYRILLIC CAPITAL LETTER DZE' \\
    't' & 'CYRILLIC CAPITAL LETTER TE' \\
    'T' & 'CYRILLIC CAPITAL LETTER TE' \\
    'u' & 'LATIN LETTER SMALL CAPITAL U' \\
    'U' & 'ARMENIAN CAPITAL LETTER SEH' \\
    'v' & 'CYRILLIC SMALL LETTER IZHITSA' \\
    'V' & 'TIFINAGH LETTER YADH' \\
    'w' & 'CYRILLIC SMALL LETTER WE' \\
    'W' & 'CYRILLIC CAPITAL LETTER WE' \\
    'x' & 'CYRILLIC SMALL LETTER HA' \\
    'X' & 'CYRILLIC CAPITAL LETTER HA' \\
    'y' & 'CYRILLIC SMALL LETTER U' \\
    'Y' & 'CYRILLIC CAPITAL LETTER STRAIGHT U' \\
    'z' & 'LATIN LETTER SMALL CAPITAL Z' \\
    'Z' & 'CHEROKEE LETTER NO'
  \end{tabular}
\end{table}

\end{document}